\documentclass[aps,prl,twocolumn,noshowpacs]{revtex4-1}
\usepackage{graphicx,epsfig, subfigure}
\usepackage{amsbsy}
\usepackage{amssymb}
\usepackage{amsmath}
\graphicspath{{figures//}}
\usepackage{pgffor} 

\date{\today}

\begin{document}

\newcommand{\eqnref}[1]{Eq.~\ref{#1}}
\newcommand{\figref}[2][]{Fig.~\ref{#2}#1}
\newcommand{\citeref}[1]{Ref. \cite{#1}}

\newcommand{\Sperp}{S}
\newcommand{\Q}{Q}

\title{
Characterizing the energy gap and demonstrating an adiabatic quench \\in an interacting spin system
}
\author{T.M. Hoang, M. Anquez, M.J. Boguslawski, H.M. Bharath, B.A. Robbins, and M.S. Chapman
}

\affiliation{School of Physics, Georgia Institute of Technology,
  Atlanta, GA 30332-0430}

\begin{abstract}
Spontaneous symmetry breaking occurs in a physical system whenever the ground state does not share the symmetry of the underlying theory, e.g., the Hamiltonian \cite{GoldstoneField1961, Sachdev2011}. 
It gives rise to massless Nambu-Goldstone modes \cite{Nambu1960, GoldstoneField1961} and massive Anderson-Higgs modes \cite{AndersonPR1963, HiggsPRL1964}. These modes provide a fundamental understanding of matter in the Universe and appear as collective phase/amplitude excitations of an order parameter in a many-body system.
The amplitude excitation plays a crucial role in determining the critical exponents governing  universal non-equilibrium dynamics in the Kibble-Zurek mechanism (KZM).
Here, we characterize the amplitude excitations in a spin-1 condensate and measure their energy gap for different phases of the quantum phase transition \cite{ChangNP2005, ZhangPRA2005}. At the quantum critical point of the transition, finite size effects lead to a non-zero gap. Our measurements are consistent with this prediction, and furthermore, we demonstrate an adiabatic quench through the phase transition, which is forbidden at the mean field level. 
This work paves the way toward generating entanglement through an adiabatic phase transition \cite{ZhangPRL2013}.
\end{abstract}

\maketitle


The amplitude mode and phase mode describe two distinct excitation degrees of freedom of a complex scalar field $\psi=A e^{i\phi}$ appearing in many quantum systems such as the order parameter of the Ginzburg-Laudau superconducting phase transition \cite{Landau1950} and the two-component quantum field of the Nambu-Goldstone-Anderson-Higgs matter field model \cite{Nambu1960, GoldstoneField1961, AndersonPR1963, HiggsPRL1964}. 
In a spin-1 condensate, the transverse spin component plays the role of an order parameter in the quantum phase transition (QPT) with $\textbf{S}_\bot$ being zero in polar (P) phase and  nonzero in broken axisymmetry (BA) phase (\figref[a]{Fig:Concept}).  
Representing the transverse spin vector as a complex number, $\textbf{S}_\bot = S_x + i S_y$, with the real and imaginary parts being expectation values of spin-1 operators, the amplitude mode corresponds to the amplitude oscillation of $\textbf{S}_\bot$.

\begin{figure}[h!]
		\includegraphics[scale=0.9]{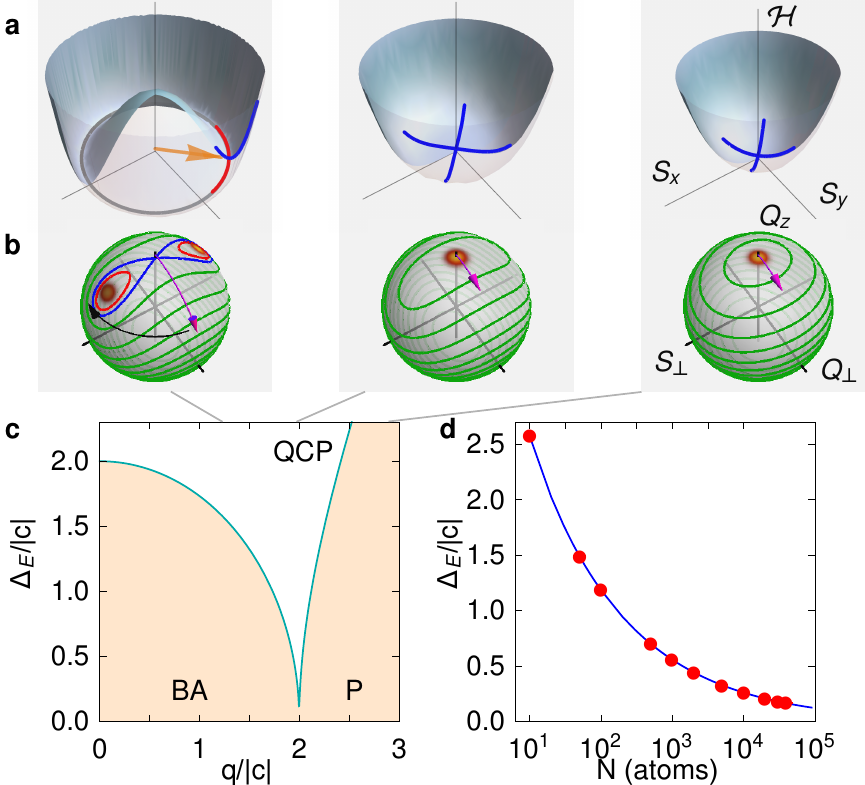}				
	\caption{  
	\textbf{a}, Spin energy in the BA phase ($q/|c|<2$), at the QCP ($q/|c|=2$), and in the P phase ($q/|c|>2$). In the P phase, there are two gapped modes (blue lines) along the radial direction about the GS. In the BA phase, the GS occupies a minimum energy ring (gray circle) with one gapped mode along the radial direction (blue line) and one NG mode (red line) in the azimuthal direction. 
	\textbf{b}, The GS on the $\{\Sperp_\bot ,  \Q_{\bot } , Q_z \}$ unit spheres is represented by a red shaded region. 
	Coherent orbiting (phase winding) dynamics are represented by red (green) curves and the blue curve is the separatrix. 
	The magenta (black) arrow represents the RF (microwave) pulse used for the  initial state preparation (Supplementary Information).
	\textbf{c}, The energy gap for 40,000 atoms (cyan curve) is calculated from the eigenvalues of the quantum Hamiltonian (Supplementary Information). 
	\textbf{d}, The energy gap at the QCP (blue solid line) calculated from the eigenvalues of the quantum Hamiltonian matches the GS oscillation frequencies from  simulations (circle markers). 
	(\textit{Color online}) }
	\label{Fig:Concept}
\end{figure}

The amplitude mode can be studied in different spinor phases
by tuning the relative strength of the quadratic Zeeman energy per particle $q \propto B^2$ and spin interaction energy $c$ of the condensate \cite{StengerNature1998} by varying the magnetic field strength $B$ (\figref{Fig:Concept}). 
In the polar phase, both the spinor energy $\mathcal{H}$ and the ground state (GS) spin vector have SO($2$) rotational symmetry about the vertical axis (\figref[a]{Fig:Concept}), and there are two degenerate collective amplitude modes along the radial directions about the GS located at the bottom of the parabolic bowl. 
These amplitude excitations are gapped modes, which vary both the amplitude of $\textbf{S}_\bot$ and the energy $\mathcal{H}$. In a second quantized picture, they correspond to pairwise excitations from $|m_F=0\rangle$ to $|m_F=\pm 1\rangle$ Zeeman spin states. 

In the BA phase, the spinor energy $\mathcal{H}$ acquires a Mexican-hat shape with the GS occupying the minimal energy ring of radius $\sqrt{4c^2-q^2}/(2|c|)$. 
The GS spin vector (orange arrow in \figref{Fig:Concept}) spontaneously breaks the SO($2$) symmetry and acquires a definite direction \cite{SadlerNature2006}.
This broken symmetry induces a massless Nambu-Goldstone (NG) mode in which it costs no energy for the spin vector to rotate about the vertical axis. Recently, the magnetic dipolar interaction was used to open a gap in the NG mode by breaking the rotational symmetry of the spin interaction \cite{MartiArxiv2014}. In our condensate, the NG phase mode remains gapless because the conserved and zero magnetization ($\langle S_z\rangle=0$) suppresses the magnetic dipolar interaction. The other excitation, the amplitude mode, manifests itself as an amplitude oscillation of the transverse spin in the radial direction. This amplitude mode is similar to the massive mode in the Goldstone model \cite{GoldstoneField1961}. 

In this work, we measure the amplitude modes in spin-1 Bose-Einstein condensates (BEC) through measurements of very low amplitude excitations from the ground state.  
The results show an quantitative agreement with gapped excitation theory \cite{MurataPRA2007, Lamacraft2007, ZhangPRL2013} and provide a new platform to probe the amplitude excitation, which plays a crucial role in the KZM in spinor condensates. 
Although in the thermodynamic limit the amplitude mode energy gap goes to zero at the quantum critical point (QCP), a small size-dependent gap persists for finite size systems \cite{ZhangPRL2013}. 
The measurements of the energy gap near the QCP are challenging, however our results are consistent with a small non-zero gap. Furthermore, by using a very slow optimized magnetic field ramp, we demonstrate an adiabatic quench across the QCP. Such adiabatic quenches in finite sized systems underlie proposals for generating massively entangled spin states including Dicke states \cite{ZhangPRL2013} and are fundamental to the ideas of adiabatic quantum computation \cite{FarhiScience2001}.


The experiments use a tightly confined $^{87}$Rb BEC with $N=40,000$ atoms in optical traps such that spin domain formation is energetically suppressed. The Hamiltonian describing this spin system in a bias magnetic field $B$ along the $z$-axis is
\cite{HoPRL1998,OhmiJPSJ1998,LawPRL1998,HamleyNature2012}:
\begin{eqnarray}
	\hat{H}=\tilde{c}\hat{S}^2 - q (\hat{Q}_z - N) /2
	\label{Hamiltonian}
\end{eqnarray}
where $\hat{S}^2$ is the total collective spin-1 operator and $\hat{Q}_z$ is proportional to the spin-1 quadrupole moment, $\hat{Q}_{zz}$. The coefficient $\tilde{c}$ is the collisional spin interaction energy per particle integrated over the condensate and quadratic Zeeman energy per particle $q=q_z B^2$ with $q_z=72$ Hz/$\mathrm{G}^2$ (hereafter, $h=1$). The  longitudinal magnetization $ \langle \hat{S}_z \rangle $ is a constant of the motion ($=0$ for these experiments); hence the first order linear Zeeman energy $p \hat{S}_z$ with $p \propto B $ can be ignored.
The spin-1 coherent states  can be represented on the surface of a unit sphere shown in \figref[b]{Fig:Concept} with axes  $ \{\Sperp_\bot ,  \Q_{\bot } , Q_z \}$ where the expectation value of transverse spin is $S_\bot^2 = S_x^2+S_y^2 $, $Q_{\bot}$ is the transverse off-diagonal nematic moment,  $Q_\bot^2 =  Q_{xz}^2 + Q_{yz}^2$, and $Q_z = 2\rho_0-1$ where $\rho_0$ is the fractional population in the $|F=1,m_F=0\rangle$ state. In this representation, the coherent dynamics evolve along the constant energy contours of $ \mathcal{H} = \frac{1}{2}cS_\bot^2 - \frac{1}{2}q (Q_z - 1) $ where $c=2N\tilde{c}$ \cite{ChangNP2005, ZhangPRA2005} (red and green orbits in \figref[b]{Fig:Concept}).

In the mean field (large atom number) limit, quantum fluctuations can be ignored and the wavefunction for each spin state, $m_F=0,\pm 1$, can be represented as a complex vector with components, $\psi_{0,\pm 1} = \sqrt{\rho_{0,\pm 1}} \exp \theta_{0,\pm 1}$. Using Bogoliubov analysis \cite{MurataPRA2007} and mean field theory \cite{ZhangPRA2005}, the energy gap of the amplitude mode in the P phase and the BA phase in the long wavelength limit correspond to the oscillation frequency of small excitations in $\rho_0$ from the GS 
\begin{small}
\begin{eqnarray}
	\Delta_{P} \equiv f_P =  2\sqrt{q(q+2c)} ,~ 
	\Delta_{BA} \equiv  f_{BA} = 2\sqrt{c^2-q^2/4} 
	\label{Eqn:gapenergy}
\end{eqnarray}
\end{small}here the energy gap is $\Delta_E$ ($\equiv \Delta_P$ and $\Delta_{BA}$) and coherent oscillation frequency is $f$ ($\equiv f_P$ and $f_{BA}$). Although these relations show a vanishing gap at the QCP, quantum fluctuations due to finite atom number size effects result in a non-zero gap. In the  quantum theory, the energy gap can be exactly calculated from the eigenenergy values of the Hamiltonian in \eqnref{Hamiltonian} (Supplementary Information). \figref[c]{Fig:Concept} shows the energy gap between the GS and first excited state with a small nonzero gap at QCP as a result of a finite atom number. \figref[d]{Fig:Concept} shows the relation of energy gap at the QCP and atom numbers for condensates ranging from $10^1-10^5$ atoms which scales as $\Delta_E \propto N^{-1/3}$ \cite{ZhangPRL2013}. The energy gap curve compares well to the oscillation frequencies of GS spinor population $\rho_0$ obtained from quantum simulations (Supplementary Information) for a broad range of atom numbers (red marker in \figref[d]{Fig:Concept}). 
The equivalence relation between the energy gap and the coherent oscillation frequency in \eqnref{Eqn:gapenergy} is a general statement connecting the amplitude modes to the observable dynamics and is key to this study.

\begin{figure*}[t!]
	\begin{minipage}{\textwidth}
		\includegraphics[scale=0.9]{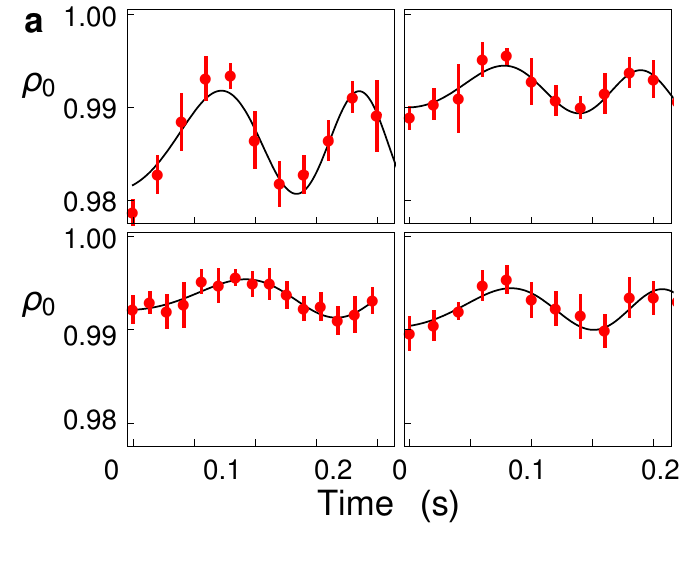}	
		\includegraphics[scale=0.9]{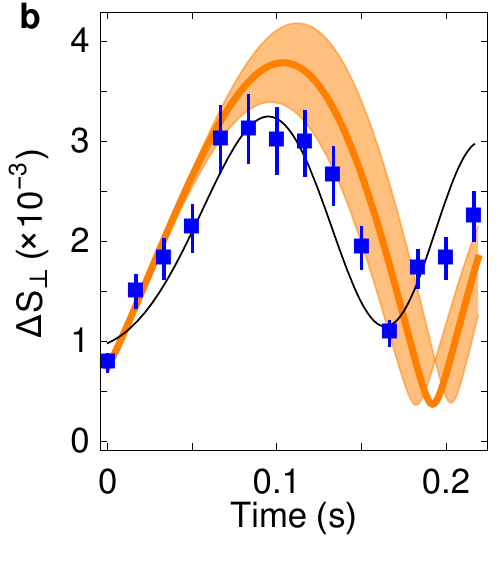}	
		\includegraphics[scale=0.9]{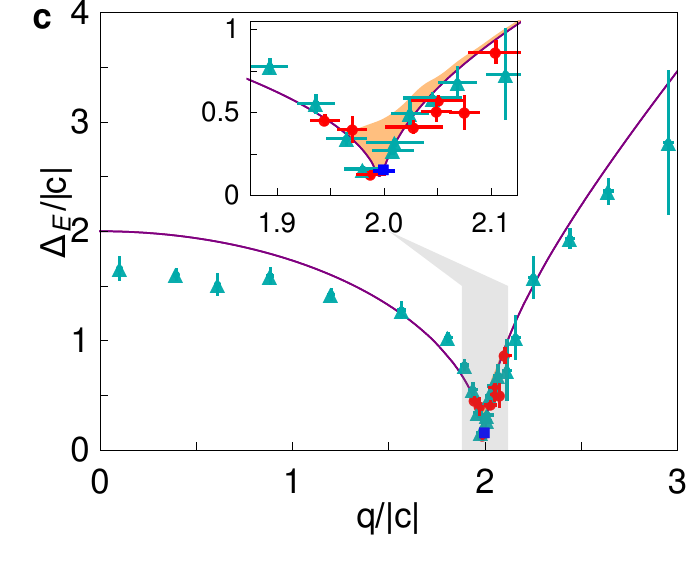}
	\end{minipage}
	\caption{\textbf{Energy gap measurements.} 
	\textbf{a}, Coherent oscillation data (circle markers) obtained at $q/|c|=2.03$. In clockwise order, the oscillation amplitude decreases as the initial state is initialized closer to the GS. Each data point is an average of 10 measurements and the data are fit to a sinusoidal function with a varying frequency (solid line)
	\textbf{b}, The time evolution of $\Delta S_\bot$ data (square markers) at the QCP ($q/|c|=2$)  are fit to a sinusoidal function (solid line). Each data point is the noise of 45 measurements. The corresponding simulation is represented by an orange curve with the shaded region being $q/|c| = 2 \pm 0.005$.
	\textbf{c}, The energy gap $\Delta_E$ for different $q/|c|$ values are obtained from the frequency fits of coherent oscillation data. Circle (triangle) markers are obtained from an average of 10 (or 3) measurements of $\rho_0$ coherent oscillation, and square marker is the frequency fit of $\Delta S_\bot$ dynamics.
The theoretical energy gap is represented by the purple curve.
The inset plot shows the region around the QCP with the shaded orange region being the energy gap for an initially imperfect GS (see text). (\textit{Color online.})
	}
	\label{Fig:EnergyGapDirect}
\end{figure*}

\textbf{Energy gap measurement.}
To characterize the energy gap $\Delta_E$, we measure coherent dynamics for states initialized close to the GS (\figref[b]{Fig:Concept}) for different values of $q/|c|$ ranging from $0.1$ to $3$ and fit the measurements to sinusoidal functions to determine the oscillation frequencies (Supplementary Information). 
For each $q/|c|$ value, several measurements of the population $\rho_0$ are made for a series of initial states approaching the GS as illustrated in \figref[a]{Fig:EnergyGapDirect}. The GS population $\rho_{0,GS}$ can be obtained by minimizing the spinor energy (Supplementary Information) \cite{ZhangPRA2005}
\begin{eqnarray}
\rho_{0,GS}=1~~(\mathrm{P})~,~  \rho_{0,GS}=1/2+q/(4|c|) ~~(\mathrm{BA})
\label{Eqn:rhogs}
\end{eqnarray}
The oscillation amplitude of $\rho_0$ has a lower limit given by the Heisenberg standard quantum limit ($\textrm{SQL}=N^{-1/2}$)  projected onto the $\rho_0$-axis ($\propto Q_{z}$-axis in \figref[b]{Fig:Concept}) \cite{HamleyNature2012}; hence the best estimate of the energy gap is obtained from the measurement with the lowest observable oscillation amplitude.  An alternate method to determine the energy gap for states centered on the pole is to measure the oscillations of the transverse spin fluctuations, $\Delta S_\bot$.  Although this method requires much more data because the signal is in the  fluctuations instead of the mean value, it provides higher contrast for states localized at the pole.  Measurements obtained with this technique at the QCP are shown in \figref[b]{Fig:EnergyGapDirect} for a state prepared in the polar GS (Supplementary Information). 

The results of the energy gap measurements are shown in \figref[c]{Fig:EnergyGapDirect} for both methods.  
Overall, the measurements  capture the characteristics of energy gap predicted by gapped excitation theory for a spin-1 BEC \cite{MurataPRA2007, Lamacraft2007, ZhangPRL2013}.
In the P phase, the energy gap data show a good agreement with the theoretical prediction within the uncertainty of the measurements.  In the BA phase, the measured gap data are also in reasonable agreement with the theory, however the measured values are 20\% lower  than the theory for the smallest values of $q/|c|<1$. This is possibly a result of small violations of the single mode approximation or the presence of a small thermal fraction, both of which would be more significant in this spin interaction-dominated regime. 
In a study of an antiferromagnetic condensate, using an initial state ($\rho_0=0.5$) prepared far away from the antiferromagnetic GS ($\rho_{0,AGS}=1$), slightly lower oscillation frequencies than the theory were also observed \cite{BlackPRL2007}; it was suggested that this resulted from excess magnetization noise from the RF pulse of the initial state preparation, however, this noise is not large enough to explain the difference in our measurements.

In the neighborhood of the QCP, the energy gap decreases dramatically.  A shown in the inset to \figref[c]{Fig:EnergyGapDirect}, the measurements are in good agreement with the theoretical predication in this region.  For measurements at $q=2|c|$, the minimum measured gap is $\Delta_E = 0.15(1)|c| $,  which is consistent with the non-zero gap  predicted by the quantum theory, $\Delta_{E,th} = 0.165 |c|$, here $c\approx -7.5(1)$~Hz (Supplementary Information).    
We point out however that there are experimental challenges to these measurements,  which can tend to over-estimate the measured value of the gap.
The initial state is prepared in the high magnetic field GS ($q/|c|=38$).  This state has symmetric $\sqrt{N}$ fluctuations in the $S_\bot, Q_\bot$ plane.  When the condensate is rapidly quenched to a lower $q/|c|$ for the energy gap measurement, this projects the condensate to slightly excited states of the final $q/|c|$ Hamiltonian. The subsequent evolution of this state will have an oscillation frequency higher than calculated gap frequency, particularly in the region $1.95 \le q/|c| \le 2.05$. We can accurately calculate this effect, and the results are indicated by the orange shaded region in \figref[c]{Fig:EnergyGapDirect}. 

A further complication in the measurement at the QCP is that the value $q/|c|$ is not truly constant during the measurement of the gap, but drifts to slightly higher values because of  a reduction of  density due to the finite lifetime of the condensate. The spin interaction energy depends on the density and atom number as $c(t) \propto n(t) \propto N(t)^{2/5}$.  For these measurements, the condensate lifetime was $1.6(1)$~s, which results in a drift of $\Delta q/|c| = 0.05$ in 100~ms in the neighborhood of the QCP. The atom loss is taken into account in the simulations, an example of which is shown in \figref[c]{Fig:EnergyGapDirect}, and the energy gap is determined by the frequency at $t=0$. Despite these challenges to the measurements near the QCP, the data indicate the presence of a non-zero gap that is of the same size as predicted by theory.

\begin{figure*}[t]
	\includegraphics[scale=0.9]{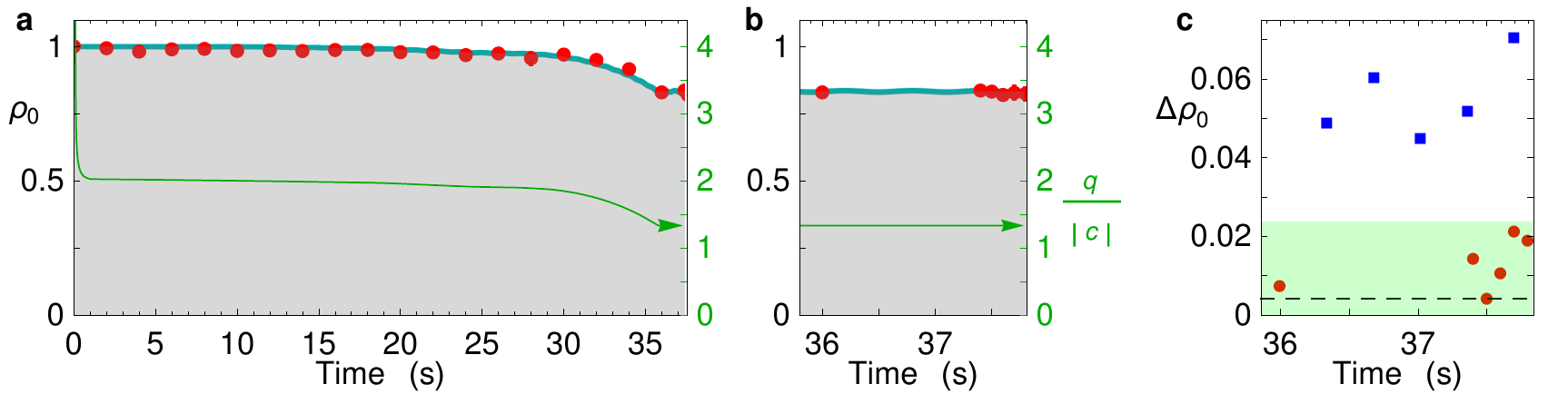}
\\	\includegraphics[scale=0.9]{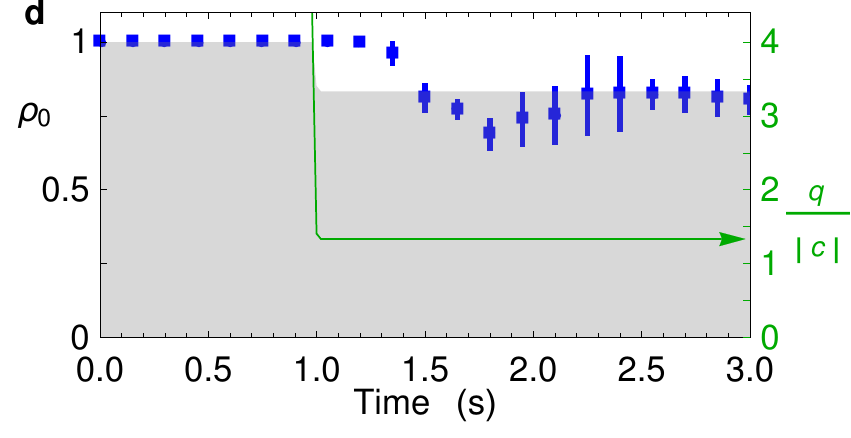}
	\includegraphics[scale=0.9]{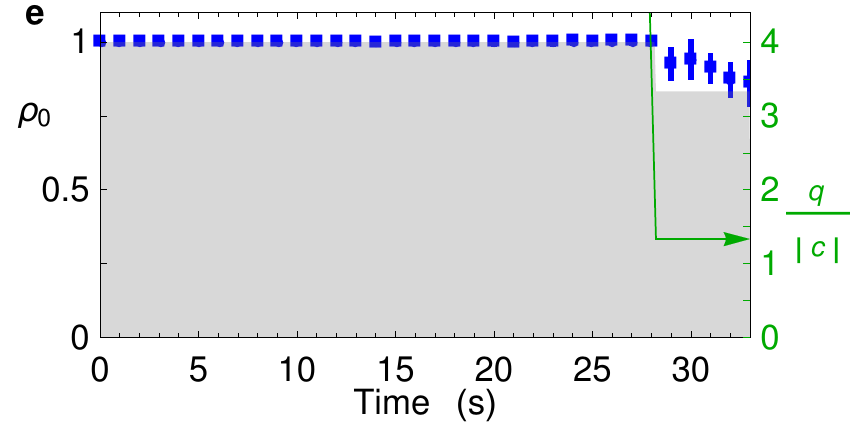}	
	\caption{\textbf{Adiabatic and nonadiabatic dynamics}. \textbf{a-c}, Adiabatic dynamics of population $\rho_0$ and the uncertainty $\Delta \rho_0$ (see texts).  
	\textbf{d-e}, Non-adiabatic dynamics of population $\rho_0$ of $1$ s and $28$ s linear ramp from $q/|c|=140 \rightarrow 1.33$. 
	Red circle (blue square) markers are the adiabatic (nonadiabatic) measurements, gray shaded regions are the theoretical $\rho_{0,GS}$ values, green arrow lines represent the ramp of $q/|c|$ (vertical axis on the right), and cyan curves represent the dynamics simulation of $\rho_0$ with the corresponding $q/|c|$ ramp.
	Each adiabatic data point is an average of 3 measurements and 
each non-adiabatic data point is an average of 15 measurements. (\textit{Color online}).
	}
	\label{Fig:adiabatic}
\end{figure*}

The $ \langle \hat{S}_z \rangle =0 $ spin-1 BEC below the QCP has a relativistic-type dispersion relation $E^2 (k) = (\Delta_E/2)^2 + c^2_s k^2$, in which the energy gap $\Delta_E/2$ is equivalent to the rest mass energy of a quasiparticle, $m_s$, and the spin wave velocity plays the role of speed light, $c_s \equiv \sqrt{q-|c|}$ \cite{MurataPRA2007, Lamacraft2007}.  Our experiments are in the long wavelength limit in which the wave vector approaches zero, $\textbf{k} \rightarrow 0$. 
The effective action of the system is identical to the theoretical description of superfluid-insulator transition in Bose-Hubbard model and hence the phase transition belongs to the same universality class as the XY model \cite{Lamacraft2007}. 
The Higgs mode manifest as a collective excitation has been observed in the superfluid/Mott insulator transition \cite{EndresNature2012} as an amplitude fluctuation of a complex order parameter, in the XY model of antiferromagnetic materials as an amplitude fluctuation of spin vector \cite{Ruegg2008}, and in superconducting systems \cite{Sooryakumar1980, Sherman2015, Pekker2015}. In a similar vein, the amplitude mode of a spinor BEC in the proximity of the QCP can be regarded as a Higgs mode associated with the amplitude fluctuation of the transverse spin. 
This Higgs mode is different from the Anderson-Higgs mechanism, in which a massless gauge field in combination with the spontaneous symmetry breaking can generate a massive boson \cite{HiggsPRL1964, AndersonPR1963}.

\textbf{Adiabatic quantum phase transition.}

In the thermodynamic ($N\rightarrow\infty$) limit, the vanishing gap at the QPT prohibits adiabatic crossing between phases  and gives rise to excitations characterized by the Kibble-Zurek mechanism (KZM).  However, the opening of the gap at the QCP due to finite size effects make it possible,  in principle, to cross the QCP adiabatically using a carefully tailored ramp from $q\gg 2|c|\rightarrow q<2|c|$, while remaining the ground state of the Hamiltonian. Recently, adiabaticity in  sodium spin-1 condensates has been studied \cite{JiangPRA2014}; however, these experiments were performed using  condensates with non-zero longitudinal magnetization ($ \langle \hat{S}_z \rangle \neq 0$) that do not have a QCP. Here, we focus on the very challenging case of small energy gaps at the quantum critical point.

Due to the small size of the gap, the ramp in $q$ needs to be very slow in the region of $2|c|$ in order to maintain adiabaticity. To allow longer ramps, we employed a single focus dipole trap in which the condensate lifetime is 15-19 s. To determine the optimal ramp, we performed simulations using measured values of the trap lifetime, the atom number, and the spin interaction energy. The ramp is determined from a piece-wise optimization of the adiabatic parameter $\frac{\mathrm{d}\Delta_E}{\mathrm{d}t}\frac{1}{\Delta_E^2}$ and includes the effects of atom loss on $c$ (Supplementary Information). The simulations show that it is possible to adiabatically cross the phase transition in $\sim$35~s starting with a condensate initially containing 40,000 atoms.  

The experiment starts with atoms at the GS in the polar phase at high magnetic field, $q/|c|=140$. Then, the magnetic field is ramped through the QCP to $q/|c|=1.33$ in 35~s along the trajectory represented by the green line in \figref[a]{Fig:adiabatic}.  The measured evolution of the population $\rho_0$ is  shown in same graph and compared to that predicted by the simulation.  The data show excellent agreement with the theoretical values for the evolving ground state population $\rho_{0,GS}$ (\eqnref{Eqn:rhogs}), which provides a strong indication of adiabitity.  

There are about 9000 atoms remaining after the adiabatic ramp. The theoretical value of the ground state population and uncertainty is $\rho_{0,GS}= 0.833 \pm 0.004$, where the uncertainty is the SQL for 9000 atoms, projected onto the $\rho_0$ axis ($\propto Q_z$-axis in \figref[b]{Fig:Concept}), (Supplemental Information).
Immediately after the adiabatic ramp ($t\approx 36$~s), the measured mean population and fluctuations are $\rho_0 = 0.830 \pm 0.007$, which are very close to the theoretical values,  and further indicate adiabiticity.
Following the adiabatic ramp, the ratio $q/|c|=1.33$ is held constant for $2$~seconds to verify that the system remains in the GS. As shown in \figref[b]{Fig:adiabatic}, the mean value of $\rho_0$ stays close to the theoretical value $\rho_{0,GS}$. In \figref[b]{Fig:adiabatic}, the uncertainty $\Delta \rho_0$ is plotted.  Although  the measurements of $\Delta \rho_0$ (red circle markers) trend above the theoretical SQL  (dashed line) after holding, atom loss increases fluctuations in the spin populations (assuming uncorrelated losses) to the level shown in the green shaded region (Supplementary Information).

For comparison, in \figref[d-e]{Fig:adiabatic} we show data from non-adiabatic ramps from $q/|c|=140 \rightarrow 1.33$. In \figref[d]{Fig:adiabatic}, a 1~s linear ramp is used, while in \figref[e]{Fig:adiabatic}, a $28$~s ramp is used.   In both cases, the spin population $\rho_0$ does not follow the theoretical GS population during the ramp and the fluctuations $\Delta \rho_0$ grow dramatically. The fluctuations at the end of the $28$~s ramp are compared with those from the adiabatic ramp in \figref[c]{Fig:adiabatic} (shown as blue square markers), and it is clear that the non-adiabaticity gives rise to increased fluctuations.

Adiabatically crossing the  QPT in a spin-1 zero magnetization condensate is predicted to generate massively entangled spin states \cite{ZhangPRL2013}. Broadly speaking, this is an example of the fundamental principle underlying  adiabatic quantum computing, in which the initial simple ground state, by tuning the Hamiltonian adiabatically through a QCP,  transforms into a highly entangled final ground state of the final Hamiltonian that is a solution to a computation problem \cite{Farhi01}. In our case, under ideal conditions, the final state would correspond to a ring-shaped GS in the BA phase as shown in \figref{Fig:Concept}{a}. For a ramp to $q=0$, the final state is predicted to be the Dicke state $|S=N, S_z=0\rangle$. In this study, we stop the adiabatic ramp at $q/|c|=1.33$. The entanglement of the GS at this $q/|c|$ can be calculated as in Ref. \cite{ZhangPRL2013}
$
	\xi = \frac{(\langle \hat{S}_x^2 \rangle+\langle \hat{S}_y^2 \rangle)N}{1+4\langle (\Delta \hat{S}_z)^2 \rangle  N^2}
$.
The uncertainty in transverse magnetization is $\langle \hat{S}_x^2 \rangle+\langle \hat{S}_y^2 \rangle \approx 1-(2\rho_{0}-1)^2$. In the ideal case, the longitudinal magnetization is zero and conserved $\langle (\Delta \hat{S}_z)^2\rangle\rightarrow 0$, and the expected entanglement is $\xi=0.56 N$ or roughly $5,000$ atoms are entangled out of $9,000$ atoms  at the end of the adiabatic ramp. However, the atom loss induces noise in the magnetization, $\Delta S_z \approx 0.5\%$, in our experiment. This small magnetization noise reduces the entanglement to $\xi < 1$ atom.

In summary, we have explored the energy gap in small spin-1 condensates. The energy gap measurements show evidence of a nonzero gap at the QCP arising from finite size effects and using a carefully tailored slow ramp of the Hamiltonian parameters, we have adiabatically crossed the QCP with no apparent excitation of the system. 
In future work, we hope to study these effects for different system sizes and to preserve and measure the entanglement of the system.  We hope that this work stimulates similar investigations in related many-body systems, and in particular, we anticipate that the results of this study could directly inform investigations in double-well Bose-Josephson junction systems, (psuedo) spin-1/2 interacting systems \cite{Milburn1997}, and the Lipkin-Meshkov-Glick (LMG) model \cite{Solinas2008}, which share a similar Hamiltonian.

\textbf{Acknowledgments}

We would like to thank C.D. Hamley and B.J. Land for their early contributions. TMH would like to thank Prof. K. Wiesenfeld and Prof. F. Robicheaux for useful discussions. We acknowledge support from the National Science Foundation Grant PHY--1208828.

{\bf Author Contributions}  
T.M.H.,  and M.S.C. jointly conceived the study. T.M.H performed the experiment and analyzed the data.    M.A., and B.A.R. assisted with the data taking.  T.M.H developed essential theory and carried out
the simulations. All authors contributed to writing the manuscript.  M.S.C supervised the work.


\bibliography{EnergyAdiabaticRefs}

\onecolumngrid
\newpage
\section{Characterizing the energy gap and demonstrating an adiabatic quench in an interacting spin system: Supplementary Information}

\section{Experimental setup} 
The experiment is carried out using small condensates of $40,000$ atoms in the $F=1$ hyperfine ground state of $^{87}$Rb. In the energy gap experiment, atoms are confined in a spherical optical dipole force trap with trap frequencies $\sim2\pi\times 140~\mathrm{Hz}$, formed by crossing the focus of a $10.6~\mu$m wavelength laser with a $850$~nm wavelength laser. This tight confinement ensures that the condensate is well described by the single mode approximation (SMA), such that the spin dynamics can be considered separately from the spatial dynamics\cite{LawPRL1998,HoPRL1998,OhmiJPSJ1998}. The spin interaction energy $c \approx -7.5(1)$~Hz and trap lifetime is $\approx 1.6(1)$~seconds. 

\textbf{Imaging protocol.}
The spin populations of the condensate are measured by releasing the trap and allowing the atoms to expand in a Stern-Gerlach magnetic field gradient to separate the $m_F$ spin components. The atoms are probed for 200~$\mu s$ with three pairs of counter-propagating orthogonal laser beams, and the fluorescence signal collected by a CCD camera is used to determine the number of atoms in each spin component.

\section{Energy gap measurement}
\begin{figure}[h!]
        \includegraphics[scale=0.5]{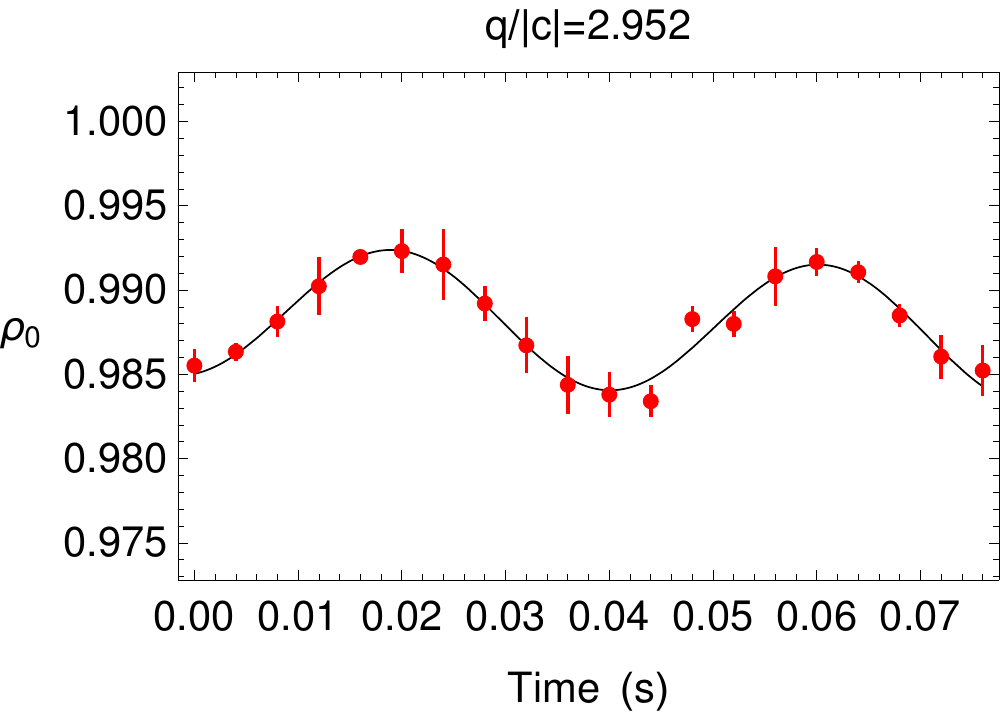}
        \includegraphics[scale=0.5]{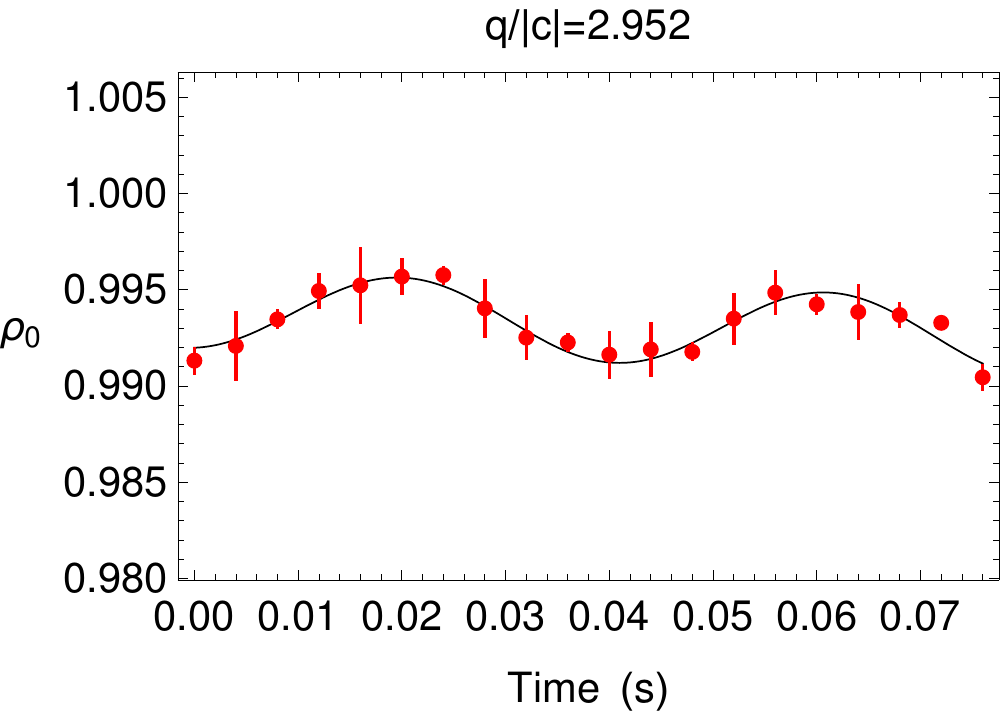}        
\caption{Coherent oscillation data (circle markers) for different $q/|c|$ values are fitted to sinusoidal functions. For each value $q/|c|$, several measurements of the population $\rho_0$ are made for a series of initial states approaching GS.  (\textit{Color online.})}
\label{Fig:AllCohrentData}
\end{figure}

In our experiment, the condensate is prepared at a high magnetic field ($q/|c|\sim 38$). This state has symmetric $\sqrt{N}$ fluctuations in the $S_\bot, Q_\bot$ plane set by the Heisenberg uncertainty limit \cite{HamleyNature2012}.
The system is subsequently quenched to a lower field in 2~ms. 
An rf pulse (transition between $|m_F=0\rangle \leftrightarrow |m_F=\pm 1\rangle$) is applied to prepare the system at $\rho_{0,GS}$ value, and a subsequent microwave pulse (transition between $|F=1,m_F=0\rangle \leftrightarrow |F=2, m_F=0\rangle $) \cite{HamleyNature2012} is applied to rotate the spinor phase $\theta_s=\theta_+ + \theta_- - 2\theta_0$ to the GS position. 
After the initial state preparation, the condensate is allowed to evolve along the energy contours, as seen on the spheres in the main paper Fig. 1b. Coherent dynamics are observed through time evolution of either the population $\rho_0$ or transverse spin component $S_\bot$. \figref{Fig:AllCohrentData} shows all the measurements of coherent oscillations for different $q/|c|$ values.

\textbf{Measuring $S_\bot$.}
Note that a $\pi/2$-rf pulse can be used to rotate $S_x$ into the $S_z$ measurement axis. Details of this protocol are described in Ref. \cite{HamleyNature2012}. As quantum states evolve along their respective energy contours, the projection of the Heisenberg uncertainty limit onto the $S_\bot$-axis oscillates and energy gap is measured through this oscillation frequency.  Since we are unable to track the Larmor phase of the spin vector due to its fast dynamics, the Larmor phase is considered to be uniformly distributed  in the spin  space $S_x S_y$. Therefore, the measurement of the standard deviation $\Delta S_x$ is equivalent to measuring $\Delta S_\bot$. 

In the main paper Fig. 2b, the condensate is prepared at the polar GS at a high magnetic field ($q/|c|\sim 38$). While the mean-field GS ($|m_F=0\rangle$ state) remains unchanged as the system is quenched to the QCP for $S_\bot$ measurements, the initial $\sqrt{N}$ quantum fluctuation projects the condensate to slightly excited states of the  Hamiltonian at the QCP. Therefore, the state used in $S_\bot$ measurements is a close approximation of the actual GS the QCP limited by the quantum noise.
\\
\\
\foreach \x in{1, ...,105}{
        \includegraphics[scale=0.5]{figure/\x.pdf}
}

\newpage
\textbf{Sinusoidal fitting.}
To extract the oscillation frequency, the data are fitted to a sinusoidal function of the form $\rho_0(t)=\rho_{00} + A \cos(2\pi f(t) t + \phi) +k\times t$ as shown in \figref{Fig:AllCohrentData}. Fitting parameters include the initial population value, $\rho_{00}$, the oscillation amplitude, $A$, the initial phase, $\phi$, and the drift of GS population due to atom loss, $k$. Due to atom loss, the frequency is a function of time, $f(t)$. Since $f_0 \equiv \Delta_E(0)$ and $f(t)\equiv \Delta_E(t)$, the frequency $f(t)=f_0 \frac{\Delta_E(t)}{\Delta_E(0)}$ where $f_0$ is the oscillation frequency at $t=0$.  The energy gap $\Delta(t)$ is calculated using the Hamiltonian in \eqnref{Eqn:MatrixH} with spin interaction energy depending on the number of atom $c(t) = c(0) \times (N(t)/N)^{2/5} $. The oscillation frequency of the $\Delta S_\bot$ dynamics is extracted with a similar fit function.

\begin{figure}[h!]
		\includegraphics[scale=1]{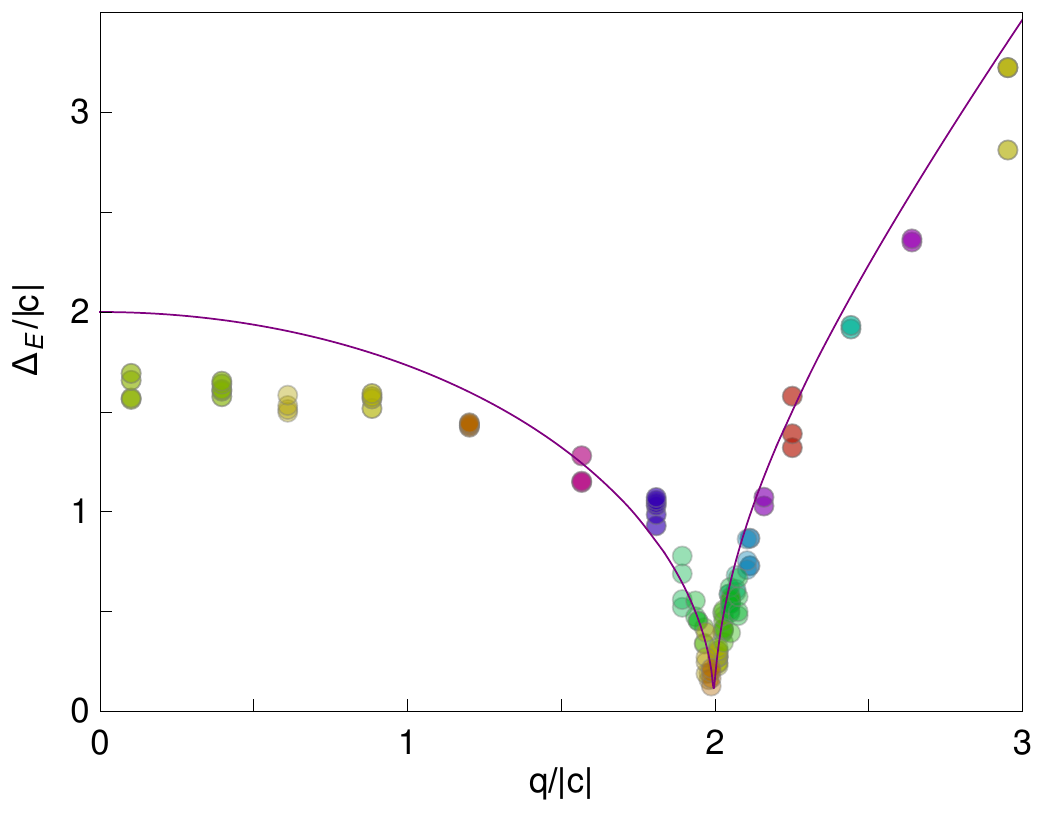}			
	\caption{Markers represent the energy gap $\Delta_E$ for different $q/|c|$ values which are obtained from the frequency fits of coherent oscillation raw data. Theoretical energy gap is represented by purple curve
. (\textit{Color online.})}
	\label{Fig:EnergyGapDirectAll}
\end{figure}

\textbf{Energy gap raw data.} The energy gap of amplitude modes is obtained by fitting data to the sinusoidal functions.
\figref{Fig:EnergyGapDirectAll} show the energy gap plot summarized from all the frequency fits. The measurements with the lowest oscillation amplitude which provides a reliable fits corresponding to the closest measurement of the energy as shown in the main paper Fig. 2c.

\section{Adiabatic QPT}
In the adiabatic QPT experiment, atoms are confined in the focus of the $10.6~\mu$m wavelength laser with the trap lifetime of $15-19$~s with spin interaction energy $c \approx -2.1$~Hz. Although the Thomas-Fermi radii of the condensate in longitudinal direction in this trap is larger than the spin healing length, the spin domains is unlikely to be formed since the adiabaticity maintains the condensate at the GS leaving no extra energy for domain formation. The single-mode approximation theory is still capable of describing the adiabatic process.

In our experiment, the condensate is prepared in a high magnetic field ($q/|c| = 140$), then subsequently a final value using an optimal ramp. An optimized ramp of the magnetic field is produced piecewise through an iterative procedure using a semi-classical simulation \cite{GervingNatureCom2012, HoangPRL2013}, in particular, we divide the whole ramp into 100 linearly small pieces. 
To verify the adiabaticity, one needs to compare the state after each linear ramp with the theoretical GS. 
Since the atom number $N$, spin interaction energy $c$, and quadratic Zeeman energy $q$ vary after each ramp, it is very time consuming to numerically solve for the exact GS from the quantum Hamiltonian in \eqnref{Eqn:MatrixH}. Instead, we use an adiabatic condition such that the uncertainty of quantum states maintains below the Heisenberg uncertainty limit with SQL$= N^{-1/2}$.
The shortest ramp time that satisfies this adiabatic condition is chosen and this procedure continues until the final $q/|c|$ value is reached. 

\textbf{Atom fluctuation}.
We assuming the atom fluctuation is equal for all spin components, $m_F=0,\pm 1$. If the atom number fluctuation is $\Delta N$, the uncertainty of number atoms in $m_F=0$ are $\Delta N/\sqrt{3}$~atom which translates into the uncertainty $\Delta \rho_{0,0} = \Delta N/(N\sqrt{3})$. 

The theoretical SQL$_{\rho_0}$ with the contribution from the number fluctuation becomes
\begin{eqnarray}
	\mathrm{SQL}_{\rho_0,\Delta N} = \sqrt{\mathrm{SQL}^2_{\rho_0} + (\Delta \rho_{0,0})^2}  \nonumber
\end{eqnarray}
During a 2~s period, the fluctuation of number atom is about 300~atoms with the mean number of atom is 9000~atom, which yields SQL$_{\rho_0} \approx 0.004$ and SQL$_{\rho_0,\Delta N} \approx  0.022$ as seen in the main paper Fig. 3c.

\section{Quantum energy gap calculation} 
The gapped excitation $\Delta_E$ can be obtained by computing the eigenvalues of the quantum Hamiltonian in the Fock basis  $| N_1, N_0, N_{-1} \rangle$, with $N_i$ atoms in the $|m_F=i\rangle$ state \cite{LawPRL1998, MiasPRA2008, GervingNatureCom2012, ZhangPRL2013},

\begin{small}
	\begin{eqnarray}
		\mathcal{H}_{k, k'} &=& \Big(  2 \tilde{c} k(2(N-2 k)-1)+2qk \Big) \delta_{k, k'+1} \notag \\
	&& + 2\tilde{c} \Big(
	(k'+1)\sqrt{(N-2 k')(N-2 k' -1)} \delta_{k, k'+1}  \notag \\
	&& + k \sqrt{(N-2k'+1)(N-2k'+2)}\delta_{k, k'-1}
	\Big)
		\label{Eqn:MatrixH}
	\end{eqnarray}
\end{small}
here $\delta_{k,k'}$ is the Kronecker delta function. With conservation of the total atom number $N=N_1+N_0+N_{-1}$ and zero magnetization $M=N_1-N_{-1}=0$, the Fock basis can be represented using $N$ and a variable $k$, which counts the number of pairs of atoms in the $m_F=\pm 1$ states. The energy gap is the excitation energy $\Delta_E$ between the lowest and first excited eigenstates. 

\section{Mean-field spinor energy}
From the main paper Eq. 1, the mean-field spinor energy can be written as
\begin{eqnarray}
	\mathcal{H} &=& \frac{c}{2} S^2_\bot - \frac{q}{2} (Q_{z} -1) 
	\label{Eqn:MeanfieldE}
\end{eqnarray}
Since the spinor dynamics is constrained on the surface of the spin-nematic sphere (main paper Fig. 1b), we have the following relationship \cite{HoangPRL2013}
\begin{eqnarray}
	S^2_\bot + Q^2_\bot + Q^2_z = 1
	\label{Eqn:SQQz}
\end{eqnarray}
At the GS, the expectation value of quadrupole operator $Q_\bot = 0$. Fixing the value $Q_\bot = 0$ in \eqnref{Eqn:SQQz}, we can expand the spinor energy in \eqnref{Eqn:MeanfieldE} around the GS as a function of transverse spin $S_\bot$,
\begin{eqnarray}
	\mathcal{H} &=& \frac{c}{2} S^2_\bot - \frac{q}{2} (\sqrt{1- S^2_\bot} -1) 
	\label{Eqn:MeanfieldESbot}
\end{eqnarray}
here the expectation value $S_\bot =  \sqrt{S^2_x + S^2_y}$. Minimizing \eqnref{Eqn:MeanfieldESbot} we obtain the GS expectation value $S_\bot=0$ in the P phase and $S_\bot = \sqrt{4c^2-q^2}/(2|c|)$ in the BA phase. The values $\rho_{0,GS}$ in the main paper Eq. 3 can be calculated from these $S_\bot$ values.
The Hamiltonian \eqnref{Eqn:MeanfieldESbot} has a global continuous SO($2$) symmetry reflected on the shape of spinor energy $\mathcal{H}$ in the main paper Fig. 1a.

\begin{figure}[t!]
        \includegraphics[scale=1]{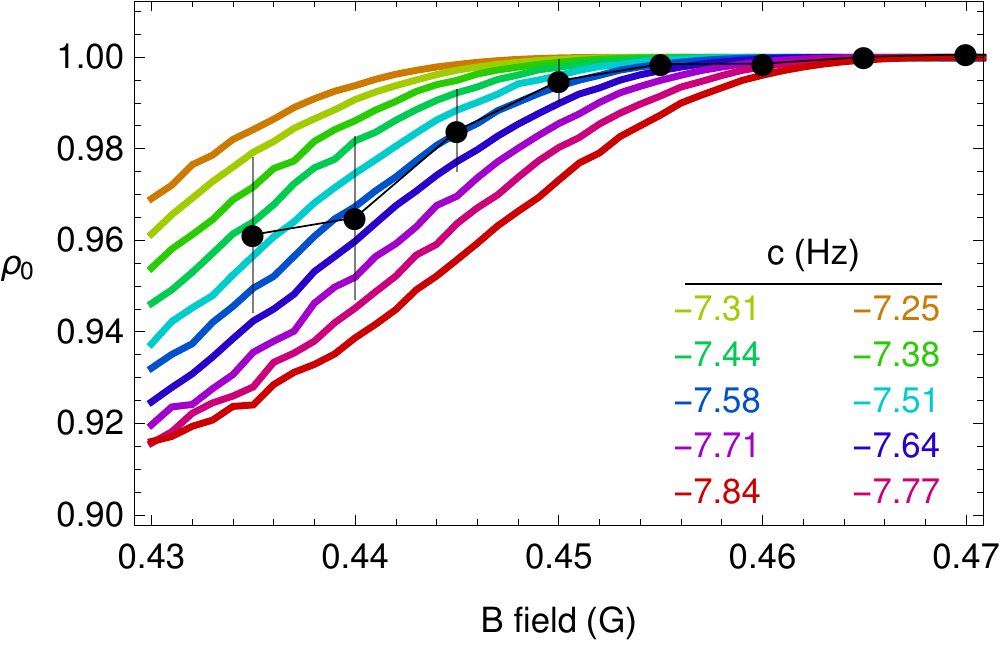}        
\caption{\textbf{An example of a spin interaction energy measurement}. The spin interaction energy, $c$, is determined by measuring population $\rho_0$ (black markers) after 150~ms time of evolution for different magnetic field values, $B$. 
Simulation results for different spin interaction energy values, $c$, are represented by the different color curves with corresponding $c$ values labeled in the same colors. (\textit{Color online.})}
\label{Fig:spininteraction}
\end{figure}

\section{Spin interaction energy}
Spin interaction energy is defined as $c=2N\tilde{c}$ with $\tilde{c} \propto N^{-3/5}$ \cite{Pu1999}, therefore, $c \propto N^{2/5}$.
If $q/|c| > 2$, the population dynamics $\rho_0$ for a condensate initialized in $m_F=0$ state remains equal to $1$ regardless of the dynamic evolution, $\rho_0(t)=1$. 
To search for the critical point $q/|c|=2$, we measure the population $\rho_0$ after a short evolution time (for instance $150$~ms) for different magnetic field values with $q = q_z B^2$ (similar to an experiment carried in Ref. \cite{ChangNP2005}). 
The magnetic field value such that the population $\rho_0$ drops slightly below 1 corresponds to $q/|c|=2$, and the spin interaction energy is calculated as $c=-q_z B^2/2$.
We can even compare the data $\rho_0$ to simulations of different $c$ values to determine $c$ with $\sim 1$\% uncertainty as shown in \figref{Fig:spininteraction}.

\section{Simulation tools}
The details of simulation method are well described in our previous works \cite{GervingNatureCom2012, HoangPRL2013}.

\bibliographystyle{nature}

\end{document}